\documentclass[twocolumn,showpacs,nofootinbib,floatfix,preprintnumbers,amsmath,amssymb,prd]{revtex4}

\usepackage{latexsym}
\usepackage{amsmath}
\usepackage{amsbsy}
\usepackage{amssymb}
\usepackage{epsfig}
\usepackage{graphicx}
\usepackage{dcolumn}
\usepackage{graphicx}
\usepackage{bm}
\usepackage{natbib}

\begin{document}

\title{Generalized Lemaitre-Tolman-Bondi solutions with Pressure}

\author{P. D. Lasky}
	\email{paul.lasky@sci.monash.edu.au}, 

\author{A. W. C. Lun}
	\email{anthony.lun@sci.monash.edu.au}

\affiliation{Centre for Stellar and Planetary Astrophysics\\
		School of Mathematical Sciences, Monash University\\
		Wellington Rd, Melbourne 3800, Australia}

		\begin{abstract}
			Utilizing the ADM equations, we derive a metric and reduced field equations describing a general, spherically symmetric perfect fluid.  The metric describes both the interior perfect fluid region and exterior vacuum Schwarzschild spacetime in a single coordinate patch.  The exterior spacetime is in generalized Painleve-Gullstrand coordinates which is an infinite class of coordinate systems.  In the static limit the system reduces to a Tolman-Oppenheimer-Volkoff equation on the interior with the exterior in Schwarzschild coordinates.  We show the coordinate transformation for the non-static cases to comoving coordinates, where the metric is seen to be a direct generalization of the Lemaitre-Tolman-Bondi spacetime to include pressures.  	
		\end{abstract}
		
		         %%%%%%%%%%%%%%%%%%%%%%%%%%%%%%%%%%%%%%%%%%%%%%%%%%%
			%%			Physics and Astronomy Classification Scheme                                                                    %%
			%%			04.20.Ex: 		Initial Value Problem, Existence and Uniqueness of Solutions             %%
			%%			04.20.Dw:		Singularities and Cosmic Censorship                                                        %%
			%%			04.70.Bw:		Classical Black Holes                                                                                    %%
			%%			04.20.Jb:		Exact Solutions                                                                                               %%
                            %%%%%%%%%%%%%%%%%%%%%%%%%%%%%%%%%%%%%%%%%%%%%%%%%%%
		\maketitle
		
\section{INTRODUCTION}
The Lemaitre-Tolman-Bondi (LTB) class of solutions of Einstein's field equations describing spherically symmetric, inhomogeneous dust  \cite{lemaitre33,tolman34,bondi47} are well known.  However, despite many years of work, the generalization to a general perfect fluid\footnote{The definition of perfect fluid we utilize in this article follows that of \cite{hawking73,misner73,stephani03}} source is not known.  There are a number of exact solutions with specific matter distributions or geometries, for example the Tolman-Oppenheimer-Volkoff (TOV) static models, the FRW models with homogeneous pressure and energy density, and the list goes on.    

Much work has been done on the collapse of self-similar perfect fluids (see for e.g. \cite{cahill71,ori90} and \cite{carr01} for a more recent review).  The field equations for simple equations of state reduce to systems of ordinary differential equations and can therefore be analysed in great detail using various methods of dynamical systems.  Giambo {\it et. al.} \cite{giambo03,giambo04} recently reduced the Einstein field equations for a perfect fluid with barotropic equation of state to a single, second order quasi-linear differential equation on the metric coefficients.  This work was achieved using area-radial coordinates, and enabled the analysis of naked singularities in the collapse of fluids of this type.  

One specific purpose of exact solutions is to describe either a stellar or galactic system.  In particular, the equations of hydrostatic equilibrium describes the pressure required to support these bodies from collapsing under their own gravitational field.  Furthermore, the non-static cases are interesting in that they describe the collapse of these systems, and hence the possible formation of black holes.  

However, many known matter solutions only describe the region that contains the matter.  The exterior region is considered as an additional extra, the matching of which provides boundary conditions for the collapse process (see for e.g. \cite{misner73}).  However, recently \cite{adler05} expressed a metric that describes both interior and exterior regions of a collapsing body as a single solution of the field equations.  This alleviates the requirement for the matching schemes at the interface, a process which is sometimes difficult due to the matching of two separate coordinate systems.

In \cite{adler05}, the interior region of the spacetime is a marginally bound ball of homogeneous dust, and the exterior spacetime is the Schwarzschild spacetime in Painleve-Gullstrand coordinates \cite{painleve21,gullstrand22}.  In \cite{mine06} we generalized this model to non-marginally bound inhomogeneous dust for the interior, with the Scwarzschild spacetime in generalized Painleve-Gullstrand coordinate system for the exterior region.  The coordinate system for the dust region of the spacetime is such that shell-crossing singularities appear as simple fluid shock waves \cite{nolan03}.

In this paper, we generalize \cite{mine06} to include pressure in the form of a perfect fluid.  The interior region is first derived as a generalization of the coordinates used in \cite{mine06}, and we further show the coordinate transformation to put this into comoving coordinates.  In this way, the solution is shown to be a generalization of the LTB spacetime to include pressures.  

By taking the static limit of our coordinates we find the system reduces to the TOV equation for hydrostatic equilibrium.  This solution matches smoothly onto exterior Schwarzschild coordinates.  Considering the minimum radius for a static star is larger than the apparent horizon, this implies the static system is everywhere regular in these coordinates.  However, if we consider the Schwarzschild coordinates for the collapsing case, then at some point through the evolution, the surface of the star will be inside the event horizon.  Therefore, the spacetime will cease to be everywhere regular.  However, the generalized Painleve-Gullstrand coordinates which are the exterior for the non-static cases are everywhere regular, and therefore the entire collapse process is everywhere well defined.      

By considering only the interior, perfect fluid region, and reversing the direction of the time coordinate the solution is a cosmological model describing the expansion of the universe.  

The article is set-up as follows.  In section \ref{reducedEFEs} we describe the method for deriving the reduced Einstein field equations.  Section \ref{IBconditions} analyses the initial and boundary conditions required to solve this set of equations.  The coordinate transformation to put the metric into a generalized version of the LTB coordinates is given in \ref{GTB}, and the apparent horizon is briefly discussed in \ref{ApparentHorizon}.  We look at the independence of the pressure on the tidal forces for the entire spacetime in \ref{TidalForces} and finally reduce the system to known solutions in section \ref{SpecificSolutions}.

Geometrized units are employed throughout whereby $c=G=1$.  Greek indices run from $0\ldots3$, Latin from $1\ldots3$ and we follow index conventions of \cite{misner73}.

\section{Reduced EFEs}\label{reducedEFEs}
The metric representing a spherically symmetric spacetime can be expressed in $3+1$ form, without loss of generality, as
\begin{align}
ds^{2}=-\alpha^{2}dt^{2}+\frac{1}{1+E}\left(\beta dt+dr\right)^{2}+r^{2}d\Omega^{2},\label{genmetric}
\end{align}
where $\alpha(t,r)>0$ is the lapse function, $\beta(t,r)$ is the radial component of the shift vector, $E(t,r)>-1$ and $d\Omega^{2}:=~d\theta^{2}+\sin^{2}\theta d\phi^{2}$.  

The energy-momentum tensor for a perfect fluid \cite{hawking73,misner73,stephani03} is given by\footnote{For more detailed discussion of relativistic hydrodynamic equations, refer the review article, \cite{font03}.  In particular, one can further split $\rho$ into a rest-mass density component and specific energy density component.} 
\begin{align}
T^{\mu\nu}=\left(\rho+P\right)n^{\mu}n^{\nu}+Pg^{\mu\nu},
\end{align}
which is related to the Einstein tensor via $G_{\mu\nu}=8\pi T_{\mu\nu}$.  Here, $\rho$ is the energy density, $P$ is the pressure and $n^{\mu}$ is the vector field tangent to the fluid, which is timelike and normalized
\begin{align}
n^{\alpha}n_{\alpha}=-1.
\end{align}  
Further demanding this satisfies Frobenius' theorem
\begin{align}
n_{[\mu}\nabla_{\nu}n_{\sigma]}=0,\label{frobenius}
\end{align}
where square brackets denote antisymmetrization, implies the normal vector is hypersurface forming.  The normal being timelike implies the hypersurfaces are spacelike.  A particular solution to equation (\ref{frobenius}) is
\begin{align}
n_{\mu}=-\alpha\nabla_{\mu}t,
\end{align}
where $t$ is the temporal coordinate which implies the normal one-form can be written in component form as
\begin{align}
n_{\mu}=\left[-\alpha,0,0,0\right].
\end{align}

The $3+1$ formalism allows for great simplification in spherically symmetric spacetimes.  In particular, consider any trace-free, symmetric, spatial two-form, $W_{ij}$ say.  Component-wise, spherical symmetry implies ${W_{\theta}}^{\theta}=~{W_{\phi}}^{\phi}$.  Furthermore, the tensor being spatial implies ${W_{t}}^{t}=0$.  Now, as the tensor is trace-free, the radial component must be related to the angular components according to ${W_{\theta}}^{\theta}=-2{W_{r}}^{r}$.  Therefore, one can write
\begin{align}
W_{ij}=w(t,r)P_{ij},  \label{eqn5}
\end{align}
where $w(t,r)$ is the distinct eigenvalue of $W_{ij}$, and
\begin{align}
{P_{i}}^{j}:=\left(\begin{array}{ccc}
-2 & 0 & 0\\
0 & 1 & 0\\
0 & 0 & 1
\end{array}\right).\label{eqn6}
\end{align}
Many of the $3+1$ variables can be expressed in terms of their eigenvalues, analogously with equations (\ref{eqn5}) and (\ref{eqn6})\footnote{The terms defined here are expressed in terms of their metric coefficients in the appendix.};  
\begin{itemize}
	\item The trace-free extrinsic curvature
\begin{align}
A_{ij}:=K_{ij}-\frac{1}{3}\perp_{ij}K:=a(t,r)P_{ij}.
\end{align}
	\item The trace-free three-Riemann tensor
\begin{align}
{^{3}Q}_{ij}:={^{3}R}_{ij}-\frac{1}{3}\perp_{ij}{^{3}R}:=q(t,r)P_{ij}.
\end{align}
	\item The trace-free Hessian of the lapse function
\begin{align}
\frac{1}{\alpha}D_{i}D_{j}\alpha-\frac{1}{3\alpha}\perp_{ij}D^{k}D_{k}\alpha:=\epsilon(t,r)P_{ij}.
\end{align}  
\end{itemize}
Here, $K_{ij}$ and $K$ are respectively the extrinsic curvature and its trace, $\perp_{ij}$ is the three-metric on spatial hypersurfaces, ${^{3}R}_{ij}$ and ${^{3}R}$ are the three-Ricci tensor and scalar respectively, and $D_{i}$ is the unique metric connection associated with $\perp_{ij}$.  

The conservation of energy-momentum and ADM equations can now be expressed as scalar equations rather than complicated tensorial relations.  The conservation equations come from the twice contracted Bianchi identities which can be expressed in terms of the energy momentum tensor
\begin{align}
\nabla_{\alpha}{T^{\alpha}}_{\mu}=0.
\end{align}  
By decomposing this onto and orthogonal to the spacelike hypersurfaces, one can derive an equation which relates the extrinsic curvature to the comoving derivative of the energy density
\begin{align}
\mathcal{L}_{n}\rho=\left(\rho+P\right)K,\label{cont}
\end{align}
where $\mathcal{L}_{n}$ denotes Lie differentiation with respect to the normal vector field.  The other equation coming from the Bianchi identities is Euler's equation, which reduces to a single component relating the pressure gradient to the lapse function
\begin{align}
\frac{\partial P}{\partial r}=-\frac{\left(\rho+P\right)}{\alpha}\frac{\partial\alpha}{\partial r}.\label{Eulera}
\end{align}      
An interesting aspect of equations (\ref{cont}) and (\ref{Eulera}) is that the equation of state $P=-\rho$, implies both $\rho$ and $P$ are constants.  This case is equivalent to a vacuum solution with cosmological constant, and while this is interesting itself we shall exclude it from the remainder of the analysis.   

The ADM system consists of ten equations separated into four constraints which are satisfied on all spacelike hypersurfaces, and six evolution equations.  In spherical symmetry, these equations reduce to just two constraint and two evolution equations.  The two constraint equations are the Hamiltonian constraint
\begin{align}
{^{3}R}+\frac{2}{3}K^{2}-6a^{2}=16\pi\rho,\label{Ham}
\end{align}
and the momentum constraint
\begin{align}
\frac{\partial}{\partial r}\left(ar^{3}\right)=\frac{-r^{3}}{3}\frac{\partial K}{\partial r}.\label{momentum1}
\end{align}
After some algebra, the evolution equations can be shown to be 
\begin{align}
2\mathcal{L}_{n}K-\frac{1}{2}{^{3}R}-K^{2}-9a^{2}+\frac{2}{\alpha}D^{k}D_{k}\alpha=&24\pi P,\label{EV1}\\
\mathcal{L}_{n}a-aK+\epsilon-q=&0. \label{EV2}
\end{align}

Now, we would like to utilize the ADM equations to express the metric in terms of physical variables.  By writing out the momentum constraint (\ref{momentum1}) in terms of the metric coefficients (see appendix), a first order differential equation on $E$ and $\alpha$ results
\begin{align}
\frac{-1}{1+E}\mathcal{L}_{n}E=\frac{2\beta}{\alpha^{2}}\frac{\partial\alpha}{\partial r},\label{LieU}
\end{align}
where $\mathcal{L}_{n}$ operating on a scalar is
\begin{align}
\mathcal{L}_{n}=\frac{1}{\alpha}\frac{\partial}{\partial t}-\frac{\beta}{\alpha}\frac{\partial}{\partial r}.
\end{align}

In general, Einstein's field equations are second order in the metric coefficients.  However, it is noted in equation (\ref{LieU}) that all second order derivative terms cancelled one another to leave a first order differential equation.  While this was the case for the momentum constraint, it does not occur for the remainder of the equations.  However, by combining linear combinations of the remaining equations, we can search for combinations of terms whereby higher order derivative terms vanish.  For instance, the combination ${^{3}R}+12q$ gives a zeroth order expression in $E$.  Therefore, adding equation (\ref{EV1}) to six times equation (\ref{EV2}) gives an algebraic expression for $E$ in terms of first order derivatives of other metric coefficients,
\begin{align}
-E=2r\left(1+E\right)\frac{\partial}{\partial r}\ln\alpha-8\pi Pr^{2}+2r\mathcal{L}_{n}\frac{\beta}{\alpha}-\left(\frac{\beta}{\alpha}\right)^{2}.                                  \label{U1}
\end{align}
Substituting this back through the Hamiltonian constraint (\ref{Ham}) implies the metric coefficients can be related to the energy density.  This equation contains a single radial derivative of the metric coefficients and pressure term
\begin{align}
4\pi\rho r^{2}=\frac{\partial}{\partial r}\left[r^{2}\left(1+E\right)\frac{\partial}{\partial r}\ln\alpha+r^{2}\mathcal{L}_{n}\frac{\beta}{\alpha}-4\pi Pr^{3}\right].\label{muh}
\end{align}
Integration of (\ref{muh}) implies the left hand side becomes a ``mass'' function\footnote{We note that this is not a physical mass quantity, however the quantity arising in a natural way suggests to the authors it is of some importance.  Furthermore, in the dust limit this becomes the familiar mass of the LTB dust solutions, and in the vacuum limit becomes the Schwarzschild mass.}  measured at some radius
\begin{align}
M(t,r):=4\pi\int_{\sigma=0}^{r}\rho \sigma^{2}d\sigma.\label{massdef}
\end{align}
The integration of the right hand side of (\ref{muh}) requires the consideration of the boundary at $r=0$.  In particular, a physical requirement for the treatment of a stellar system is that the radial derivative of the pressure vanish at $r=~0$.  Along with Euler's equation and the regularity of $\beta$ and $E$ at $r=0$, this implies the first term on the right hand side vanishes at $r=0$.  Regularity of $\beta$, $\alpha$ and $P$ at $r=0$ further implies that the remaining two terms in the right hand side also vanish at $r=0$.  Therefore, the integration process, and the substitution of Euler's equation yields
\begin{align}
\frac{M}{r^{2}}+4\pi Pr=\mathcal{L}_{n}\frac{\beta}{\alpha}-\frac{1+E}{\left(\rho+P\right)}\frac{\partial P}{\partial r},\label{Monrsq}
\end{align}   
where $\rho\neq-P$.  

Substituting equation (\ref{Monrsq}) through equation (\ref{U1}) results in an algebraic expression for the lapse, shift and $E$
\begin{align}
E+\frac{2M}{r}=\left(\frac{\beta}{\alpha}\right)^{2}.\label{Edef}
\end{align}
While the metric being explicitly dependant on the lapse function is necessary due to Euler's equation, equation (\ref{Edef}) implies we no longer require explicit dependence on the shift function.  Instead, everywhere we substitute the shift function $\beta$, for 
\begin{align}
\beta=\alpha\sqrt{\frac{2M}{r}+E},\label{shift}
\end{align}
where we have taken only the positive root.  Consistently taking the negative root gives an expanding model rather than a contracting model.  This can be seen as changing the sign of the shift vector is equivalent to reversing the time coordinate (i.e. $t\rightarrow-t$).

Putting equation (\ref{shift}) back through equation (\ref{LieU}) we find an evolution equation for the energy function relating to the pressure gradient
\begin{align}
\frac{1}{1+E}\mathcal{L}_{n}E=2\sqrt{\frac{2M}{r}+E}\frac{1}{\rho+P}\frac{\partial P}{\partial r}. \label{LieE44}
\end{align}
Substituting (\ref{shift}) into equation (\ref{Monrsq}) we find 
\begin{align}
\mathcal{L}_{n}\sqrt{\frac{2M}{r}+E}=\frac{1+E}{\rho+P}\frac{\partial P}{\partial r}+\frac{M}{r^{2}}+4\pi Pr.\label{Liesqrt}
\end{align}
Finally, by expanding out the Lie derivative in the above equation and utilizing the evolution equation on the energy function, we can reduce this to a much simpler equation
\begin{align}
\mathcal{L}_{n}M=4\pi r^{2}P\sqrt{\frac{2M}{r}+E}.  \label{NSL}
\end{align}

Summarily, the system can be expressed as the line element along with two evolution equations
\begin{align}
ds^{2}=-\alpha^{2}&dt^{2}+\frac{\left(\alpha\sqrt{2M/r+E}\,\,dt+dr\right)^{2}}{1+E}+r^{2}d\Omega^{2},\label{metric}\\
\frac{\partial M}{\partial t}-&\alpha\sqrt{\frac{2M}{r}+E}\left(\frac{\partial M}{\partial r}+4\pi Pr^{2}\right)=0, \label{LieM}\\
\frac{\partial E}{\partial t}-\alpha&\sqrt{\frac{2M}{r}+E}\left(\frac{\partial E}{\partial r}+2\frac{1+E}{\rho+P}\frac{\partial P}{\partial r}\right)=0,\label{LieE}
\end{align}
where $\rho\neq-P$, and the lapse function satisfies Euler's equation
\begin{align}
\frac{1}{\alpha}\frac{\partial\alpha}{\partial r}=\frac{-1}{\rho+P}\frac{\partial P}{\partial r}.\label{Euler}
\end{align} 
As it stands the system is underdetermined and must be closed by providing an equation of state
\begin{align}
f\left(\rho,P\right)=0.\label{EOS}
\end{align}
Once the equation of state is prescribed, equation (\ref{Euler}) can be integrated to find the lapse in terms of the energy density, and then the coupled equations (\ref{LieM}) and (\ref{LieE}), can be solved simultaneously.    
   
We make a quick note here regarding the remaining coordinate freedom associated with the line element given in (\ref{metric}).  It is straightforward to see this is invariant under the usual rotation of the two-sphere, and a re-scaling of the time coordinate.  This is pertinent for the following sections, as one can see when the radial derivative of the pressure vanishes, equation (\ref{Euler}) implies that the lapse function is simply a function of the temporal coordinate.  However, using the freedom to re-scale this coordinate, this function can be scaled to unity.     

\section{Initial and Boundary conditions}\label{IBconditions}
\subsection{Initial Conditions}
Equations (\ref{LieM}-\ref{EOS}) are intrinsically difficult to solve analytically, in part due to the freedom associated with specifying an equation of state.  Whether solving analytically or numerically, these equations require the input of initial and boundary conditions.  As we are establishing a physical problem, it is desirable to simply prescribe two quantities on the initial hypersurface 
\begin{align}
\rho(0,r)\qquad\textrm{and}\qquad \left.\frac{\partial\rho}{\partial t}\right|_{t=0}.   
\end{align}
Through the definition for the mass (\ref{massdef}), this implies the initial mass and its time rate of change are known.  Furthermore, an equation of state implies the initial pressure and its time rate of change are known, and through Euler's equation (\ref{Euler}), the initial lapse function is known.  Equation (\ref{LieM}) can be rearranged such that
\begin{align}
E=\left\{\frac{1}{\alpha\left[4\pi r^{2}\left(\rho+P\right)\right]}\frac{\partial M}{\partial t}\right\}^{2}-\frac{2M}{r},\label{Einitial}
\end{align}
implying the energy function on the initial hypersurface is also determined.  Thus, prescribing the two quantities associated with the density on the initial hypersurface provides sufficient initial conditions.

\subsection{Boundary Conditions}
Before discussing the free boundary condition we first look at the simpler interior and exterior conditions.  

At $r=0$, all functions are required to be regular for a finite time in the evolution.  This will cease to be correct once the matter has collapsed to a point, at which point only the Schwarzschild solution remains.  Extra to the regularity condition is that the radial derivative of the pressure vanishes at $r=0$.  

There are many ways to utilize boundary conditions at the interface between the matter filled interior and the vacuum exterior.  If we were to simply let the energy density, as well as the pressure vanish at a single free-boundary, then in general, the interior and exterior do not continuously match.  This is seen by looking at equation (\ref{LieE}) and Euler's equation (\ref{Euler}), which imply the energy function $E$, and the lapse function $\alpha$, are respectively not defined unless the radial derivative of the pressure vanishes faster than $\rho+P$.  We note that simple equations of state, for example a linear equation of state, do not satisfy this condition.  We can therefore always prescribe this condition as a boundary condition for the equation of state.  Alternatively, we can define the equation of state such that the pressure drops to zero faster than the energy density.  In this way, there is a finite region consisting of dust, such that the pressure has become negligible.  Physically, this represents a sparse atmosphere around the collapsing object and is consistent with the concepts introduced in \cite{misner73} to explain phenomena such as white dwarfs.  Of course, one does not have to impose this condition, in which case the vacuum will not match continuously to the interior.  

The mathematics that has come from providing the atmosphere has prompted reasonable physics.  The observer for the spacetime is travelling with some velocity relative to the coordinates given by the shift function.  We can imagine the transition as the observer travels from the vacuum region into the matter region.  If the observer is suddenly confronted with a region of perfect fluid with non-zero pressure, then the velocity of this observer will contain a jump discontinuity.  However, if they first travel through an atmosphere, then both transitions may be made continuously.

We therefore have two free boundaries for the spacetime.  The first, denoted $r_{\partial_{1}}(t)$, is the interface defined by the point at which the pressure vanishes and energy density remains.  The second free boundary, denoted $r_{\partial_{2}}(t)>r_{\partial_{1}}(t)$, is the interface defined by the vanishing of the energy density.  In this way, our equation of state is now given by
\begin{align}
0=\left\{\begin{array}{ccc}
f\left(\rho,P\right) & \textrm{for} & r\in\left(0,r_{\partial_{1}}\right)\\
P & \textrm{for} & r\in\left[\left.r_{\partial_{1}},r_{\partial_{2}}\right)\right..
\end{array}\right.
\end{align}
The inner of the two free boundaries $r_{\partial_{1}}(t)$, can now be determined by specifying an initial radius $r_{\partial_{1}}(0)$, where $P$ goes to zero.  Demanding $\rho$ (hence $M$), and $E$ be continuous, and $P=0$, in equations (\ref{LieM}), (\ref{LieE}) and (\ref{Euler}) enables the evolution of this boundary to be determined.    

The outer free boundary, $r_{\partial_{2}}(t)$, is given such that $\rho=~P=0$ for all $r>r_{\partial_{2}}(t)$.  This implies the mass function is a constant, and furthermore, one can show that this mass in the exterior region must necessarily be equivalent to the exterior mass on the initial hypersurface.  Thus, the mass in the exterior is given by the Schwarzschild mass.  

The boundary condition on both free boundaries for the energy function is the simple demand of continuity.  That is, 
\begin{align}
\lim_{r\rightarrow r_{\partial}(t)^{-}}E(t,r)-\lim_{r\rightarrow r_{\partial}(t)^{+}}E(t,r)=0.
\end{align}
This boundary condition can then be used to determine the unique coordinate system for the Schwarzschild spacetime that describes both the interior and exterior regions in a single coordinate patch.

\section{Generalized Lemaitre-Tolman-Bondi}\label{GTB}
General inhomogeneous dust solutions are most commonly expressed in LTB \cite{lemaitre33,tolman34,bondi47} coordinates.  We therefore find the coordinate transformation such that the perfect fluid system is in coordinates that generalize the LTB model.  By letting $\left(t,r,\theta,\phi\right)\mapsto\left(T,R,\theta,\phi\right)$, such that $t=t(T)=~T$ and $r=r(T,R)$ where
\begin{align}
\left(\frac{\partial r}{\partial T}\right)^{2}=\alpha^{2}\left(\frac{2M}{r}+E\right),\label{coord}
\end{align}  
where now $\alpha$, $M$ and $E$ are all functions of $T$ and $R$, the metric becomes
\begin{align}
ds^{2}=-\alpha^{2}dT^{2}+\frac{1}{1+E}\left(\frac{\partial r}{\partial R}\right)^{2}dR^{2}+r^{2}d\Omega^{2}.
\end{align}
Furthermore, this coordinate transformation can be put through equations (\ref{LieM}) and (\ref{LieE}) to give
\begin{align}
&\frac{\partial M}{\partial T}=4\pi Pr^{2}\alpha\sqrt{\frac{2M}{r}+E},\label{masssomething}\\
\frac{\partial r}{\partial R}&\frac{\partial E}{\partial T}=2\frac{1+E}{\rho+P}\frac{\partial P}{\partial R}\alpha \sqrt{\frac{2M}{r}+E},\label{ET}
\end{align}
and Euler's equation reads
\begin{align}
\frac{1}{\alpha}\frac{\partial \alpha}{\partial R}=\frac{-1}{\rho+P}\frac{\partial P}{\partial R}.\label{TBEuler}
\end{align}
While the system in the $(t,r)$ coordinates described the black hole region of the spacetime, and the time reverse (i.e. $t\rightarrow-t$) described the white hole region, the $(T,R)$ coordinates now describe both these regions.  Furthermore, the Jacobian of transformation is zero only when $\partial r/\partial R$ vanishes.

It is apparent how the system of equations (\ref{coord}-\ref{TBEuler}) reduces to the LTB system as the pressure vanishes.  This is therefore a direct generalization of the LTB model to include inhomogeneous pressures.  Furthermore, by looking at this reduction of the system, one can appreciate the function $E$, is simply the energy function of the standard LTB model.  In this way, the cases of $E$ being positive, negative or equivalently zero correspond to unbound, bound and marginally bound models respectively.     

It is interesting to note that the mass function is still in terms of the original definition given in equation (\ref{massdef}), which now reads
\begin{align}
M(T,R):=4\pi\int_{\sigma=0}^{r(T,R)}\rho\sigma^{2}d\sigma.
\end{align}
This can be differentiated with respect to the LTB radial coordinate $R$, and rearranged to give
\begin{align}
\rho(T,R)=\frac{M^{\prime}}{4\pi r^{\prime}r^{2}},
\end{align}
where a prime denotes differentiation with respect to $R$.  This is the standard equation for the pressure-free LTB solution (see for e.g. \cite{christodoulou84,newman86}), and indicates shell-crossing singularities occur for $r^{\prime}(T,R)=0$ and shell-focussing singularities for $r(T,R)=0$.  The occurrence of these singularities is now implicitly dependant on the pressure due to equation (\ref{masssomething}).   

An interpretation of these singularities is awkward due to the choice of coordinates, implying they are exhibited simply as an infinite density.  However, in the $(t,r)$ coordinates, the singularities are exhibited as a multi-valued mass function.  This is akin to fluid shock waves (c.f. pressure-free case \cite{nolan03,mine06}).  Further analysis of these shocks will appear in a future article.

\section{Apparent Horizon}\label{ApparentHorizon}
To establish the apparent horizon for a spacetime, one must analyse both the metric as well as the equations of motion.  In particular, utilizing the unique affine parameter associated with the metric, $\lambda$, and the Euler-Lagrange equations, there are five equations governing the dynamics of all null geodesics,
\begin{align}
\mathcal{L}=&\frac{-\alpha^{2}\left(1-2M/r\right)}{1+E}\dot{t}^{2}+\frac{2\alpha\sqrt{2M/r+E}}{1+E}\dot{t}\dot{r}\nonumber\\
&+\frac{\dot{r}^{2}}{1+E}+r^{2}\dot{\theta}^{2}+r^{2}\sin^{2}\theta\dot{\phi}^{2}=0,\\
0=&\frac{d}{d\lambda}\frac{\partial\mathcal{L}}{\partial\dot{x}^{\mu}}-\frac{\partial\mathcal{L}}{\partial x^{\mu}},
\end{align}
where a dot represents differentiation with respect to $\lambda$, and $\mathcal{L}=\mathcal{L}\left(\dot{x}^{\mu},x^{\mu}\right)$ is the Lagrangian.  Spherical symmetry implies only the radial geodesics are required, and we therefore set $\dot{\theta}=\dot{\phi}=0$.  By utilizing the remaining three equations, one can solve for $\dot{t}$ and $\dot{r}$ with one arbitrary constant of integration.  We next define a null vector according to  
\begin{align}
k^{\mu}:&=\frac{dx^{\mu}}{d\lambda},
\end{align}
which is therefore everywhere tangential to the congruence of radial null geodesics.  The arbitrary constant acts to scale this vector, and it is therefore set to unity without loss of generality.  We find
\begin{align}
k^{\mu}=\sqrt{1+E}\left[\frac{1}{\alpha},\sqrt{1+E}-\sqrt{\frac{2M}{r}+E},0,0\right],
\end{align}
where the radial coordinate is now parametrized by the temporal coordinate.  It is straightforward to verify this vector is null, and the expansion of this vector field is simply it's divergence,
\begin{align}
\Theta:=\nabla_{\alpha}k^{\alpha}.
\end{align}
This is a measure of the convergence and divergence of the congruence of radial, null geodesics.  The limiting case of the converging and diverging geodesics are those given by a vanishing expansion factor, $\Theta=0$, and this defines the apparent horizon \cite{wald84}.  Evaluating this term gives
\begin{align}
\Theta=\frac{\sqrt{1+E}}{\alpha r^{2}}\frac{\partial}{\partial r}\left[\alpha r^{2}\left(\sqrt{1+E}-\sqrt{\frac{2M}{r}+E}\right)\right].
\end{align}
By setting $\Theta=0$, integrating from $r=0$ to some finite radius and using the regularity of the functions at $r=0$ implies the apparent horizon is given by the parametric equation
\begin{align}
r\left(t\right)=2M\left(t,r\left(t\right)\right).
\end{align}
This simple form shows that at the interface between the matter and vacuum regions, where the mass function simply becomes the Schwarzschild mass, the apparent horizon reduces to the familiar event horizon in the Schwarzschild spacetime.  

\section{Tidal forces}\label{TidalForces}
Ellis \cite{ellis71} showed that the tidal forces of a spherically symmetric spacetime are given by the electric conformal curvature $E_{ij}$, which arises from the decomposition of the Weyl tensor.  The electric conformal curvature is a trace-free, spatial two-tensor, and therefore has a unique eigenvalue, denoted $\lambda$.  Spherical symmetry further implies this is the only non-zero contribution from the Weyl tensor.  A constraint equation from the gravito-electromagnetic system of equations relates the divergence of the electric curvature to the gradient of the energy density
\begin{align}
D_{k}{E^{k}}_{i}=\frac{8\pi}{3}D_{i}\rho.
\end{align}
Evaluating this in terms of its eigenvalue implies the equation can be integrated, and we find the relation between $\lambda$ and $\rho$ is given by
\begin{align}
\lambda=\frac{-r}{3}\frac{\partial}{\partial r}\left(\frac{M}{r^{3}}\right).
\end{align}
It is interesting to note the tidal forces on the spacetime are explicitly independent of the pressure (this is the same result achieved in \cite{mine06}).  Furthermore, the exterior Schwarzschild spacetime is described by $M=M_{s}$, and the usual tidal force for the Schwarzschild spacetime results.

While in general these are Petrov type D solutions, the system can be reduced to a Petrov type 0 solution.  These are found when the above eigenvalue vanishes, which in turn gives an FRW interior solution.  The FRW solutions have homogeneous matter distributions, implying the pressure and density are constant along each $t$ equals constant slice.  This homogeneity, along with the spherical symmetry, in turn implies there are no tidal forces felt by any particles in the spacetime.  

\section{Analytic Solutions}\label{SpecificSolutions}
There are many methods available to us in searching for analytic solutions of equations (\ref{LieM}-\ref{Euler}).  For instance, one can employ an equation of state, solve (\ref{Euler}) and attempt to solve the remaining two coupled equations.  This process, while its motivations are purely physical, is overrun with mathematical difficulties.  For instance, choosing an arbitrary equation of state will, in general, imply the time variation equation on the mass is highly non-linear.  

We therefore make assumptions which are more directed at simplifying the mathematics, in order to grasp a better handling of the physics.  Once the mathematics is simplified, we can analyse the resulting equation of state to ascertain its physical relevance.  

We have already determined that utilizing an equation of state such that $P=0$ implies the system reduces to the LTB dust interior with generalized Painleve-Gullstrand exterior region \cite{mine06}.  Another major class of important solutions are the static solutions.

\subsection{Static Solutions}\label{TOV}
We retrieve static solutions of the field equations by removing all time dependance.  From equation (\ref{LieM}) we either derive $P=-\rho$, which is a static (anti) de-Sitter model, or
\begin{align}
E=-\frac{2M}{r}.
\end{align}
This is a statement implying that the shift vector $\beta$, is identically zero.  The metric is therefore given by
\begin{align}
ds^{2}=-\alpha^{2} dt^{2}+\frac{dr^{2}}{1-2M/r}+r^{2}d\Omega^{2},\label{TOVmetric}
\end{align}
and furthermore, equation (\ref{LieE44}) implies $E=E(r)$, which implies $M=M(r)$, and in turn $\rho=\rho(r)$.  Finally, putting this back through equation (\ref{Liesqrt}) and utilizing Euler's equation 
\begin{align}
\frac{-1}{\rho+P}\frac{dP}{dr}=\frac{M+4\pi Pr^{3}}{r^{2}\left(1-2M/r\right)}=\frac{1}{\alpha}\frac{d\alpha}{dr},
\end{align}
which is exactly the TOV equation of hydrostatic equilibrium.  

The static solutions are a special case of the equations we have derived, and all other cases (that is, with $E\neq~-~2M/r$), result in non-static systems.  While this is not a new result, it is interesting to note that letting the pressure and density vanish at some finite radius implies an exterior region is simply given by the Schwarzschild spacetime in Schwarzschild coordinates.  To see this we let $P\rightarrow0$ and $\rho\rightarrow0$ (by utilizing the atmosphere discussed in section \ref{IBconditions}), and find the solution for the lapse function is
\begin{align}
\alpha=\sqrt{1-\frac{2M_{s}}{r}}.
\end{align}
Putting this into (\ref{TOVmetric}) gives the Schwarzschild metric in Schwarzschild coordinates.  

Therefore, the natural coordinates for the exterior of the static model are the Schwarzschild coordinates.  While the generalized Painleve-Gullstrand coordinates are everywhere regular, the Schwarzschild coordinates have a coordinate singularity at a finite radius, $r=2M_{s}$.  However, the interior for the static case must necessarily have $r_{\partial}>2M_{s}$, and therefore the coordinates for the entire spacetime are still everywhere regular.  

\subsection{Non-Static Solutions}
\subsubsection{Linear Mass Equation}
The form of the mass equation (\ref{LieM}) is interesting.  Essentially, once an equation of state is prescribed, both the pressure and the lapse function can be written in terms of radial derivatives of the mass function.  We therefore ask the question: under what conditions will equation (\ref{LieE}) be a quasi-linear differential equation?  That is, with a linear radial derivative.  This question can be expressed mathematically as
\begin{align}
\alpha\left(\frac{\partial M}{\partial r}+4\pi r^{2}P\right)=k\frac{\partial M}{\partial r},
\end{align}
where $k$ is a constant of proportionality.  Rearranging, and using the definition of the mass function implies
\begin{align}
\alpha=\frac{k\rho}{\rho+P}.
\end{align}
Substituting this through Euler's equation (\ref{Euler}) implies
\begin{align}
\frac{P}{\rho\left(\rho+P\right)}\frac{\partial\rho}{\partial r}=0.
\end{align}
Therefore, equation (\ref{LieM}) is linear either in the absence of pressure (see \cite{mine06}) or if the energy density is independent of the radial coordinate.  Considering a reasonable equation of state with this latter condition implies $P$ is also independent of the radius.  These two conditions are exactly an FRW interior with vacuum exterior.  

The right hand side of Euler's equation (\ref{Euler}) now vanishes, implying the lapse is simply a function of the time coordinate.  Residual coordinate freedom can be utilized to re-scale the time coordinate such that $\alpha=1$ without loss of generality.  This implies only a linear, barotropic equation of state is valid
\begin{align}
P=\left(k-1\right)\rho.\label{BEOS}
\end{align}  

The differential equation on the mass function (\ref{LieM}) can now be expressed in terms of the density and energy functions
\begin{align}
\frac{d\rho}{dt}=\frac{3}{r}\sqrt{\frac{8\pi}{3}\rho r^{2}+E}\,\,\left(\rho+P\right).\label{FRW1}
\end{align}
Functional dependance for this equation is consistent if and only if
\begin{align}
E(t,r)=\chi(t)r^{2}.
\end{align}
These solutions are actually characterized by another aspect of the solution.  Namely, they form a subset of the {\it shear-free} solutions.  Shear-free solutions are those such that $A_{ij}=0$, which in the formalism given above is equivalent to $a(t,r)=0$.  In general, these solutions can be shown to satisfy $\sqrt{2M/r+E}=rf(t)$, for some function of integration $f(t)$.  The set of shear-free solutions we discuss differ from the shear-free solutions in \cite{stephani03} as $A_{ij}$ is a coordinate dependant quantity.

Returning to the smaller class of FRW solutions, $\chi$ and $\rho$ now satisfy a coupled system of first order ordinary differential equations
\begin{align}
\frac{d\rho}{dt}=&3\sqrt{\frac{8\pi}{3}\rho+\chi}\left(\rho+P\right),\\
\frac{d\chi}{dt}=&2\sqrt{\frac{8\pi}{3}\rho+\chi}\,\,\chi.\label{chi}
\end{align}
This system can be solved for the interior region using the equation of state (\ref{BEOS}), to give
\begin{align}
\rho^{3k}=A\chi^{2},
\end{align}
where $A$ is an integration constant.  This system can further be solved analytically when the constant $k$, is specified.  Furthermore, at some finite radius on the initial hypersurface, the density and pressure will be zero.  In this region, equation (\ref{chi}) can be calculated to find the unique exterior region of the spacetime that matches onto the FRW interior.     

\subsubsection{Spatially Quasi-Flat Solutions}
We can also search for solutions that contain some sense of spatial flatness.  Letting the three-Ricci tensor vanish implies spacelike hypersurfaces are flat.  This condition implies the energy function necessarily vanishes, which in turns gives $P=P(t)$.  This is therefore equivalent to the FRW solutions discussed above. 

Alternatively, we can let the three-Ricci scalar ${^{3}R}$, vanish.  This implies $E=f(t)/r$, where $f(t)$ is a function of integration.  Without loss of generality, we can utilize residual coordinate freedom to re-scale the time coordinate such that $f=1$.  The energy function is therefore given by 
\begin{align}
E=\frac{1}{r}.
\end{align}  
Therefore, the energy plays a similar role to the mass function as they are both represented via terms which are inversely proportional to the radial coordinate.  By rewriting equation (\ref{LieE}), and using Euler's equation, we find
\begin{align}
\frac{1}{2r\left(r+1\right)}=\frac{1}{\rho+P}\frac{\partial P}{\partial r}=\frac{-1}{\alpha}\frac{\partial\alpha}{\partial r}.
\end{align}
Without an equation of state, this equation can be integrated for the lapse function, yielding
\begin{align}
\alpha=g(t)\sqrt{1+\frac{1}{r}},\label{thing1}
\end{align}
where $g(t)$ is a function of integration.  Furthermore, by just specifying $P=P(\rho)$, we find a formal integral for the energy density
\begin{align} 
\int\frac{1}{P\left(\rho\right)+\rho}\frac{dP}{d\rho}\frac{\partial\rho}{\partial r}dr=\ln\left(\frac{1}{g(t)}\sqrt{\frac{r}{r+1}}\right).\label{thing2}
\end{align}
Upon specifying an equation of state, equations (\ref{thing1}) and (\ref{thing2}) can be substituted back through the evolution equation for the mass (\ref{LieM}) to establish constraints for the arbitrary function $g(t)$.

\subsubsection{Self-Similar Solutions}
While self-similar fluid solutions have been readily treated in the literature (for a recent review see \cite{carr01}), it is a worthwhile exercise to establish how they arrive in the context of the equations derived herein.  

Spherically symmetric self-similar solutions can be put into a form where all dimensionless quantities are functions of the self-similar variable $\xi:=t/r$ \cite{cahill71}.  For instance, we define $H\left(\xi\right):=2M/r$ and $p\left(\xi\right):=4\pi Pr^{2}$.  Furthermore, $\alpha$, $E$ and $\beta$ are all dimensionless, and are therefore functions of the self-similar variable $\xi$.  This implies the density becomes
\begin{align}
8\pi\rho r^{2}=H-\frac{dH}{d\xi}\xi.
\end{align}    
Equations (\ref{LieM}), (\ref{LieE}) and (\ref{Euler}) respectively become
\begin{align}
\frac{dH}{d\xi}=&\frac{\beta\left(H+2p\right)}{1+\beta\xi},\\
\frac{1}{1+E}\frac{dE}{d\xi}=&\frac{-4\beta}{\mu\left(1+\beta\xi\right)}\left(\frac{dp}{d\xi}\xi-2p\right),\\
\frac{1}{\alpha}\frac{d\alpha}{d\xi}=&\frac{-2}{\mu\xi}\left(\frac{dp}{d\xi}\xi-2p\right),
\end{align}
where
\begin{align}
\mu\left(\xi\right):=H-\frac{dH}{d\xi}\xi+2p,
\end{align}
and
\begin{align}
\beta\left(\xi\right)=\alpha\sqrt{H+E}.
\end{align}
The system has therefore reduced to a complicated set of coupled ordinary differential equations.  Although the system is now written as ordinary differential equations, it is apparent that they are still highly coupled, and therefore difficult to solve analytically.  However, physical aspects of the solutions can be explored through methods of dynamical systems.

\section{Conclusion}
We have derived reduced field equations describing a spherically symmetric spacetime with a perfect fluid region in the interior and an exterior vacuum region.  The interior is shown to be a direct generalization of the LTB dust solutions to include inhomogeneous pressures.  A general solution to these reduced field equations is not found, and it is not obvious that one such solution exists.  Despite this, we have shown a handful of specific solutions that were already known.  It is part of the robustness of the formalism that everything from the static, TOV equations, as well as the FRW models, the self-similar solutions and many more known solutions fall naturally from the equations derived herein.  

Before contemplating exact solutions of the reduced equations, an equation of state must be prescribed.  Within the present work we have made no attempt to prescribe such an equation due to the unreliability of the present understanding of microscopic physics under extreme pressures and densities.  Rather than prescribe an equation of state, it is a common technique to use geometrical or mathematical simplifications such that the field equations are put into a form that is soluble (see \cite{stephani03} for a review).  From this point, mathematically plausible equations of state are derived.  While this method yields tractable solutions, it is still an open question as to whether the derived equations of state are physically reasonable for high density and pressure regimes.   

The above method is often used to analyse the TOV equation describing the static case.  This is because only a handful of analytic solutions are known with simple polytropic equations of state.  While we claim no new ground on this solution, we point out that it arises naturally from our formalism.  The TOV equation is the only case that has a diagonal metric, implying it is also the only case where the Schwarzschild coordinates provide a natural description for the exterior region. 

An interesting aspect of this is that the solution is always regular through the horizon.  This is obvious in the collapsing cases, whereby both the interior and exterior regions are described by coordinate systems that are everywhere regular ($r\neq0$).  However, we showed the static case reduces to the diagonal metric, with Schwarzschild coordinates in the exterior.  This system is still everywhere regular, as a static object must necessarily have fluid covering $r=2M_{s}$.  

There are many avenues for future work within the realms of this formalism.  Obviously, solving the system of derived differential equations both analytically and numerically with prescribed equations of state is a desirable research goal.  Another aim is to study the formation and evolution of both shell-crossing and shell-focussing singularities.  While this has been extensively researched for the dust models, very little is understood about the cases with the inclusion of pressures.   

The model we have described is also able to be generalized through many avenues.  Firstly, the inclusion of heat flux and anisotropic pressure terms will allow for more realistic stellar models.  With only a perfect fluid interior and Schwarzschild exterior, the model is akin to a pressure cooker.  As the system collapses, the temperature will naturally rise, however, there is no allowance within the model for the inclusion of radiation in the form of either photons or gravitational waves.  The inclusion of aniostropic terms for the interior, and generalizing the exterior to the Vaidya spacetime will allow for incoherent null radiation (see for e.g. \cite{lake81}, \cite{lake82}, \cite{hellaby94}, and references therein).  Relaxing the spherical symmetry to quasi-spherical symmetry will allow for gravitational radiation where the exterior will be a Robinson-Trautman spacetime.  One final achievable generalization of the model presented herein is to derive fluid equations for the axisymmetric case.  This will include a rotating fluid for the interior with a Kerr exterior.

\acknowledgments
All calculations were checked using the computer algebra programme Maple.  The authors wish to thank the referees for their constructive comments and useful suggestions regarding the manuscript.  One of the authors (PL) wishes to thank the Australian government for the award of an APA scholarship.

\appendix
\renewcommand{\theequation}{A\arabic{equation}}
\setcounter{equation}{0}  
\renewcommand{\thesection}{\arabic{section}}
\setcounter{section}{0}
\section*{APPENDIX}\label{App}  
Terms in the ADM equations are expressed as functions of the metric coefficients for clarity
\begin{align}
K=&\frac{1}{\alpha r^{2}}\frac{\partial}{\partial r}\left(r^{2}\beta\right)+\frac{1}{2\left(1+E\right)}\mathcal{L}_{n}E,\\
a=&\frac{-r}{3\alpha}\frac{\partial}{\partial r}\left(\frac{\beta}{r}\right)-\frac{1}{6\left(1+E\right)}\mathcal{L}_{n}E,\\
{^{3}R}=&\frac{-2}{r^{2}}\frac{\partial}{\partial r}\left(rE\right),\\
q=&\frac{r}{6}\frac{\partial}{\partial r}\left(\frac{E}{r^{2}}\right),\\
\epsilon=&\frac{-r\sqrt{1+E}}{3\alpha}\frac{\partial}{\partial r}\left(\frac{\sqrt{1+E}}{r}\frac{\partial\alpha}{\partial r}\right),\\
\frac{1}{\alpha}D^{k}D_{k}\alpha=&\frac{\sqrt{1+E}}{\alpha r^{2}}\frac{\partial}{\partial r}\left(r^{2}\sqrt{1+E}\frac{\partial\alpha}{\partial r}\right).
\end{align}

\bibliography{PressurePaper3rdSubmission14Sept06}

\end{document}